\documentclass[12pt,a4paper, captions=figureheading]{scrartcl}
\usepackage[utf8]{inputenc}
\usepackage[T1]{fontenc}
\usepackage{lmodern}
\usepackage{ragged2e}
\usepackage{pdfpages}
\setkomafont{disposition}{\normalcolor\bfseries\rmfamily}
\usepackage{todonotes}
\usepackage[affil-it]{authblk}
\usepackage{amsmath, amsfonts, amssymb, bbm, dsfont}
\usepackage{graphicx, color, enumitem, soul, verbatim}
\usepackage{geometry} 
\usepackage{tabularx}
\usepackage{appendix}
\usepackage[colorlinks=true, allcolors=blue, breaklinks=true]{hyperref}
\usepackage{makecell}
\usepackage{setspace, float, placeins, adjustbox}
\usepackage{pdflscape, multirow}
\usepackage{gensymb}
\usepackage[anythingbreaks]{breakurl}
\usepackage{subcaption}
\usepackage{tocloft}
\makeatletter
\renewcommand{\numberline}[1]{{\@cftbsnum #1\@cftasnum~}\@cftasnumb}
\makeatother
\setkomafont{caption}{\small}
\setkomafont{captionlabel}{\usekomafont{caption}}

\usepackage[style=authoryear, citetracker=true, maxcitenames=2, hyperref=true, url=true, isbn=false, doi=true, natbib=true, 
bibstyle=authoryear, maxbibnames=30, sorting=nyt, uniquename=full, uniquelist=false]{biblatex}
\usepackage{csquotes}
\newpage

\setlist{nosep}

\title{Making intellectual property rights work for climate technology transfer and innovation in developing countries}

\vspace{2cm}

	\author{Su Jung Jee$^{1,2}$\footnote{Corresponding authors: SJ: s.j.jee@sheffield.ac.uk; KH: khotte@turing.ac.uk}, Kerstin H\"otte$^{2,3*}$, Caoimhe Ring$^{4,5}$, Robert Burrell$^{4,6}$\\
		
		\vspace{1cm}
		\footnotesize{$^{1}$ Sheffield University Management School, University of Sheffield, UK}\\
		\footnotesize{$^{2}$ Institute for New Economic Thinking, University of Oxford, UK}\\
		\footnotesize{$^{3}$ The Alan Turing Institute, London, UK}\\
		\footnotesize{$^{4}$ Faculty of Law, University of Oxford, UK}\\
		\footnotesize{$^{5}$ University of Bristol Law School, UK}\\
		\footnotesize{$^{6}$ Melbourne Law School, Australia}
	}
	
		\date{\today}
		
\begin{document}
			
			\pagenumbering{roman}
			\maketitle
			
			
			\abstract{This study investigates the controversial role of Intellectual Property Rights (IPRs) in climate technology transfer and innovation in developing countries. Using a systematic literature review and expert interviews, we assess the role of IPRs on three sources of climate technology: (1) international technology transfer, (2) adaptive innovation, and (3) indigenous innovation. Our contributions are threefold. First, patents have limited impact in any of these channels, suggesting that current debates over IPRs may be directed towards the wrong targets. Second, trademarks and utility models provide incentives for climate innovation in the countries studied. Third, drawing from the results, we develop a framework to guide policy on how IPRs can work better in the broader context of climate and trade policies, outlining distinct mechanisms to support \textit{mitigation} and \textit{adaptation}. Our results indicate that market mechanisms, especially trade and demand-pull policies, should be prioritised for mitigation solutions. Adaptation differs, relying more on indigenous innovation due to local needs and low demand. Institutional mechanisms, such as finance and co-development, should be prioritised to build innovation capacities for adaptation.}

		\vspace{1cm}
		\newpage
			
			\section*{Acknowledgements} 
			The authors want to thank the attendants of the research seminar of the Institute for New Economic Thinking in Oxford for their comments. 
			This article is based on the study “Intellectual Property Rights, Climate Technology Transfer and Innovation in Developing Countries”, commissioned by the Deutsche Gesellschaft für Internationale Zusammenarbeit (GIZ) GmbH on behalf of the German Federal Ministry for Economic Cooperation and Development (BMZ). The contents of this article do not represent the official position of neither BMZ nor GIZ. The authors gratefully acknowledge valuable comments on the research received by GIZ and BMZ members. 
			\newpage
			
		\section*{Highlights}
		
		\begin{itemize}
			\item Focus on IPR restrictions in international climate policy discussions seems more distracting than helpful.
			\item Trademarks and utility models have the potential to promote climate technologies in developing countries.
			\item Global markets for \textit{mitigation} offer scope for technology transfer and adaptive innovation. 
			\item \textit{Adaptation} relies more on indigenous innovation due to the region-specific needs and lack of markets.
			\item A framework is developed to guide how IPRs can work better in the broader context of climate and trade policies at regional, national and international levels.
		\end{itemize}
		\newpage
		\pagenumbering{arabic}
		
		\begin{quote}
			\emph{``[R]emoving obstacles to knowledge sharing and technological transfer – including intellectual property constraints – is crucial for a rapid and fair renewable energy transition.''}\\ 
			\raggedleft \footnotesize
			UN Secretary-General Antonio Gueterres (May 2022)
		\end{quote}
		
		\section{Introduction}
		Climate change adaptation and mitigation technologies can help developing countries cope with existential climate risks and achieve sustainable development pathways \citep{ipcc2022climate}. The international community, and developed countries in particular, committed to providing technological support to help developing countries achieve their climate goals \citep{UNFCCC2022progress, UNFCCC2015Paris}. Nevertheless, the diffusion and use of climate technologies\footnote{See section \ref{ct and ipr} for a definition of climate technologies.} are lagging behind the needs. 
		
		
		It is controversial whether intellectual property rights (IPRs) help or hinder developing countries in meeting their technological needs \citep{athreye2023intellectual, tur2018patents, sarnoff2011patent, drahos2011six}. Critics argue that IPRs increase the costs of accessing essential climate technologies: IPRs prevent local inventors from engaging in adaptive follow-on innovation based on foreign technologies, which could be a stepping stone in sustainable development as well as technological catch-up \citep{dreyfuss2021technological, urias2020access, maskus2010differentiated}. Especially the Trade-Related Aspects of Intellectual Property Rights (TRIPS) agreement, which is an international legal agreement on minimum standards of intellectual property (IP) among World Trade Organization (WTO) member countries, has been criticised for countervailing the goals of the Paris Agreement \citep{zhou2019can}.
		
		In contrast, proponents of stringent IPRs point to their theoretical justification: IPRs enable inventors to appropriate the gains of their inventions, and thereby act as an incentive for inventors to invest in R\&D and engage in international technology transfer (ITT) \citep{lopez2009innovation, abdel2015intellectual, zhang2016trade, cockburn2016patents, spielman2016private}. Despite this theoretical disagreement, the state of the evidence on these topics is poor \citep{maskus2009intellectual, ring2021patent}.  Indeed, there is little knowledge on how domestic IP systems impact the diffusion of climate technologies \citep{ipcc2022climate}.
		
		This study contributes evidence to this debate, especially regarding climate technology transfer and innovation in developing countries by addressing the question: \textit{Are existing technology transfer mechanisms and the current TRIPS framework suitable for promoting the diffusion of climate change technologies? If so, how and when do they work? If not, what reforms can make national and international policy work better?}
  
        IPRs over impactful climate technologies are predominantly held by developed countries while developing countries bear the brunt of climate change without having contributed much to its causes and lack access to relevant technologies. Article 4.1 of the United Nations Framework Convention on Climate Change (UNFCCC) recognises this imbalance in the principle of \emph{`common but differentiated responsibilities'} (CBDR), outlining mechanisms under Articles 4(1)(c), 4(3) and 4(5) for developed nations to facilitate the transfer of climate technologies to developing countries.\footnote{\label{CBDR}\url{https://unfccc.int/resource/docs/convkp/conveng.pdf}} 
		
		Given the complex and ambiguous relationship between IPRs and technological development, the heterogeneity among climate technologies, and the poor coverage of developing countries by quantitative data, we conducted a systematic literature review and 20 in-depth interviews with IP experts from four selected countries (Bangladesh, Kenya, India, South Africa) to understand the current status and future directions. 
		The interview questions rely on a literature review on the nexus between IPRs, development, and innovation. From the review, we identify three main areas where IPRs interact with climate technology: (1) international technology transfer (e.g. trade, foreign direct investment (FDI), technological cooperation), (2) indigenous innovation (original development of solutions) and (3) adaptive innovation (marginal changes in existing technology). 
		
		The subsequent thematic analysis reveals three major insights: 
		First, IPRs currently do not play a significant role in these areas, indicating that both sides in the heated debate on IPRs and climate technology likely overstate their importance as a barrier or driver of diffusion in developing countries. We found no clear rationale for, or against, weakening IPR protection under the TRIPS Agreement, such as introducing TRIPS waivers for climate technologies.
		
		Second, although we find a low relevance of patents, our results point to the potential of trademarks as an alternative means of IPR protection in fostering indigenous and adaptive climate innovations by local entrepreneurs. Specifically, trademarks appear more accessible than patents as a useful tool for marketing and quality assurance. We also found support for using utility models (second-tier patents for less sophisticated inventions) to promote simple yet high-impact climate innovation in developing countries. 
		
		Third, based on the findings, we develop a governance framework for incorporating IPRs in the broader context of national and international climate and trade policies, especially outlining distinct support mechanisms for mitigation and adaptation and distinguishing between market- and institution-oriented policy. Institutional support, such as finance and co-development, should be prioritised for adaptation where the market is immature, but climate solutions are urgently needed. In contrast, for mitigation, markets are mature, but demand is low. We conclude that market support mechanisms (e.g. tariff reductions) and demand-pull policies (e.g. environmental regulation, reducing local deployment costs), combined with institutional support, appear more relevant than relaxing IPRs for mitigation. 
		
		The results of this study extend to related policy discussions, including those on IPRs in the context of COVID-19, industrial policy and the developmental/entrepreneurial state, with relevance to policymakers at regional, national and international levels. 
		
		The remainder is structured as follows: Section \ref{sec:background} reviews the literature on IPRs, ITT and innovation in the climate context. Sections \ref{sec:med} and \ref{sec:results} are dedicated to the fieldwork. Sections \ref{sec: discussion} and \ref{sec:discussion mitigation adaptation} discuss the policy implications. Section \ref{sec:conclusion} concludes by highlighting its limitations and future research agendas.
		
		\section{IPRs, climate technology transfer and innovation}
		\label{sec:background}
		We analyse the debate around IPRs and climate technology in developing countries through three themes: IPRs relevance to climate technologies (Section \ref{subsec:IPR and ct}), IPRs impact on ITT (Section \ref{subsec:IPR and ctt}), and their influence on innovation and development (Section \ref{subsec:CT Inno}).
		
		\subsection{IPR and climate technology}
		\label{subsec:IPR and ct}
		\subsubsection{IPRs: rationale and complex reality}\label{subsubsec:IPR ra}
		Classical economic theory views IPRs as a mechanism to address the public good nature of knowledge, allowing innovators temporary monopolies over the knowledge embodied in their creations via patents, design rights, or copyrights. Theoretically, this system incentivizes R\&D by enabling inventors to charge higher prices, ensuring a return on investment \citep{hall2014choice, lopez2009innovation, arrow1962economic}.
		
		Micro-level interactions between innovation and IPRs are more complex for several reasons. First, innovators often rely on informal IP protection like trade secrets and confidentiality agreements, with the relative importance of formal and informal mechanisms varying across sectors and technologies \citep{jee2021perceived, hall2014choice}. 
		Second, motivations for engaging in R\&D are diverse and not solely profit-driven \citep{bollinger2021innovation, phillips2015social, seelos200912, taussig1915inventors}. Many breakthroughs, especially in climate solutions, stem from public research and universities incentivised by public funding and scientific curiosity. In these areas, IPRs might be less relevant \citep{hotte2022knowledge, mazzucato2011entrepreneurial, thursby2007university}. 
		Third, seeking formal IP protection is not solely about protecting an invention. Strategic motivations include attracting finance, using patents as collateral or as a signalling device \citep{mann2018creditor, chava2017lending, hsu2008patents, long2002patent} or for defensive purposes like avoiding IP litigation \citep{macdonald2004means}. 
		
		The significance of these motivations varies across sectors, with limited robust evidence of formal IPRs impact on knowledge creation and this coming primarily from the pharmaceutical and chemical sectors \citep{granstrandinnovation}.
		
		\subsubsection{Climate technology and IPRs in developing countries}\label{ct and ipr}
		Climate technologies are technologies \emph{`that help us reduce greenhouse gases and adapt to the adverse effects of climate change'} \citep{UNFCCC2022progress}, whereby technology is \emph{`a piece of equipment, technique, practical knowledge or skills for performing a particular activity'} \citep{ipcc2000methods}. They include tangible (e.g. equipment) and intangible items (e.g. know-how), transferred within and between countries.
		
		Climate technologies are diverse, and their effectiveness in adaptation and mitigation depends on the local context, including technological and financial capacities, social acceptance, incumbent technologies, and geographic factors \citep{olawuyi2018technology}. Since the impact of IPRs is technology-specific \citep{granstrandinnovation}, a differentiated view is needed \citep{maskus2010differentiated}.
		
		Four key factors are notable for analysing the impact of IPRs. 
		First, many climate technologies, particularly renewable energy, are mature, and various competing solutions exist. Hence, concerns that IPRs limit access to these technologies may be misplaced, as relevant IPRs often have expired. 
		Second, many climate risks in developing countries require low-tech or nature-based solutions (NbS), which rarely rely on patented technologies \citep{dreyfuss2021technological, seddon2020global}. 
		Third, cutting-edge patented solutions from advanced economies may be unfit for developing countries, and low-tech and NbS offer more suitable alternatives with co-benefits for other Sustainable Development Goals (SDGs) \citep{seddon2020global}. 
		Fourth, \emph{effective} diffusion requires more than access. Absorptive capacities, including capabilities to understand, maintain, use, and eventually adapt imported technology, are needed to ensure effective and sustainable use of these technologies \citep{olawuyi2018technology}. 
		
		The relationship between IPR, innovation and diffusion is complex and involves synergies and conflicts. For example, some climate technologies are not yet mature and there might be a reason for concern about IPRs impeding ITT of emerging technologies, in the areas of clean manufacturing, aviation, agriculture, biotechnology, and synthetic fuels \citep{maskus2010differentiated}. More generally, stringent IPRs can promote R\&D investments and technology diffusion but may lead to technology monopolies, excessive pricing, and hinder adaptive innovation \citep{dreyfuss2021technological, maskus2010differentiated}. 
		
		There is also one specific issue about mitigation technologies, as they contribute to the global public good of mitigating climate change. In this, they show a second public good feature, next to the positive knowledge externality that applies to all forms of innovation. This is commonly called the ‘double externality’ of climate technologies, justifying public support for innovation and diffusion \citep{jaffe2005tale}.

		\subsection{IPR and climate technology transfer}
		\label{subsec:IPR and ctt}
		ITT is embodied in various market-based activities like trade, licensing, and FDI. 
		The impact of IPR protection on ITT varies across these activities, with previous studies offering different conclusions, depending on their field of study or empirical focus. 
		ITT also occurs through institutional routes such as development assistance, where IPRs play a minor role. However, these institutional support mechanisms can enhance absorptive capacities, facilitating market-based ITT \citep{olawuyi2018technology}.
		
		Economic studies often show a positive link between strong IPR protection and market-based ITT. For instance, \citet{cockburn2016patents} report a positive impact of stringent IPRs on the global diffusion of new drugs. \citet{montobbio2015iprs} 
        find IPRs to enhance R\&D collaborations. In agriculture, IPRs have been found to help close the yield gap between developing and developed countries \citep{spielman2016private}. Such studies suggest that firms more often transfer technologies to countries with robust IP protection, especially in technology-intensive sectors \citep{maskus1998role}.
		
	    Studies in development and international relations emphasize the importance of strong IPRs for trade and FDI. Developing countries are encouraged to enhance protection for IPRs to facilitate the importation of innovations, although the suggested level of IPR protection varies across countries and sectors. Weak IP protection in Africa has been said to hinder technology transfer, contributing to a lack of absorptive capacity \citep{olawuyi2018technology}. Adopting and implementing TRIPS standards may boost inward FDI in developing countries \citep{zhang2016trade}. Stronger IPRs in developing countries may also promote exports of new products, especially in sectors like pharmaceuticals and medical equipment \citep{abdel2015intellectual}. However, FDI in developing countries is mainly due to economic opportunities, such as market size, purchasing power, and growth potential \citep{elfakhani2015analysis, moon1994beyond}, which are often more important than IPRs.
		
		The legal literature tends to question the efficacy of IPRs such as patents for ITT in climate technologies, focusing particularly pre-market R\&D incentives \citep{ring2021patent, tur2018patents, sarnoff2011patent, drahos2011six}. Scholars argue that patents may not incentivize climate technology innovation due to the lack of demand for these solutions \citep{tur2018patents, olawuyi2018technology}. There are also longstanding debates about whether existing international IP and climate agreements establish sufficient ITT obligations for developed countries \citep{zhuang2017intellectual, rimmer2011intellectual}. Part of this discussion has focused on amendments to IP law, including restrictions on the patenting of essential climate technologies. This has led to proposals such as greater reliance on compulsory licensing and government use provisions, with these proposals drawing on experience in the public health domain and aiming to address obstacles posed by high licensing costs and patent holders’ refusals to license \citep{zhou2019can, vindigni2016rethinking}.
		
		\subsection{IPR and climate technology innovation} \label{subsec:CT Inno}
		Developing countries use climate technologies in different ways. Some climate-related challenges like transitioning to clean energy and transport are shared by countries, irrespective of their geography or development level. Many potential solutions, including renewable energy, clean transport, and carbon capture technologies, are already available, although cost-effective implementation and scaling up remain challenging \citep{iea2020a, ipcc2022climate, sovacool2008placing}. Developing countries can access these solutions through technology transfer channels like FDI, trade, or licensing. This approach is crucial in sophisticated high-tech areas such as biotechnology, clean combustion, and batteries. 
		
		In many cases, however, countries will need to adapt foreign technologies. Follow-on \textit{adaptive innovation} can address specific regional demands or enhance efficiency. For example, implementing clean technologies requires follow-on research and local adaptation, like optimizing them for the local infrastructure or climatic conditions \citep{raiser2017corporatization}. Adaptive innovation involves incremental improvements building on existing technologies. Effective adaptive innovation requires absorptive capacity, which is the ability to recognize the value of new technology, assimilate it, and apply it to commercial ends \citep{sasidharan2011foreign, hobday2000innovation, cohen1990absorptive}. Countries lacking this capability struggle to absorb or adapt foreign climate technologies \citep{li2011sources, olawuyi2018technology}. Often, successful adaptive innovation for domestic use requires building indigenous innovation capabilities. Local adaptive innovation therefore depends on a country’s development stage and other local circumstances \citep{aghion2005growth}.
		
		Beyond adapting foreign technologies, there is also a need for locally specific technologies, especially in climate change adaptation. Climate risks and solutions are often localised, despite the global scope of climate change \citep{dodman2012adapting}. Developing countries, especially those with higher temperatures, face climate threats to food security, economic growth, and public health. Technologies to address local risks are therefore essential, especially in agriculture and to fight epidemics \citep{senyagwa2022africa}. 
		
		Most climate solutions are developed in advanced economies, and their R\&D prioritises solutions to address the needs of developed rather than developing countries \citep{papaioannou2014inclusive}. Incentives to develop technologies tailored to the needs of developing countries are often absent, given low purchasing power and the presence of cultural, institutional, and economic barriers to diffusion \citep{zhuinstitutionalized}. Consequently, there is underinvestment in technologies that would address the unique challenges of developing countries. \textit{Indigenous innovation} by local entrepreneurs can help close this technology gap. Due to limited technological capacities, indigenous innovations are often low-tech solutions, utilizing unskilled labour and ensuring easy maintenance in conditions characterised by unstable electricity supply and weak supply chains \citep{ockwell2016improving, fu2011role, lam1994beyond}. 
		
		IPRs might play a role in the above forms of technological development, but may not impact the local development of NbS or simple, low-tech solutions. IPRs tend to be geared towards high-tech, expensive innovations, rather than the low-cost, low-complexity but impactful innovations produced in developing countries \citep{kapczynski2012continuum}. 
		IPRs, especially patents, are unlikely to significantly enable or inhibit ITT of low-tech solutions between developing countries. This topic has gained importance, as developing countries sometimes face similar climate risks and socioeconomic conditions, making solution sharing viable \citep{urban2018china}. Recent trends show increased indigenous innovation and technology transfer among developing and emerging economies \citep{herman2021green, corvaglia2014south}. 
		The extent to which IPRs can play any role (positive or negative) in this emerging channel of ITT needs to be established through empirical evidence.
		
		However, there is a legitimate concern that IPRs have the potential to deter developing countries from engaging in the adoption and adaptation of climate technologies. For technology adoption, IPRs may increase the costs of accessing essential climate technologies. Commercial actors in a few developed countries, including the US, Japan, and Germany, hold the majority of climate patents \citep{jee2024knowledge}. Adaptive innovation building on existing core technologies risks infringing IPRs, which hinders their creative utilization. 
		However, empirical evidence about these infringement risks being a barrier to climate technology diffusion in developing countries is lacking. 
		
\begin{figure}
			\centering
			\includegraphics[width=0.55\textwidth]{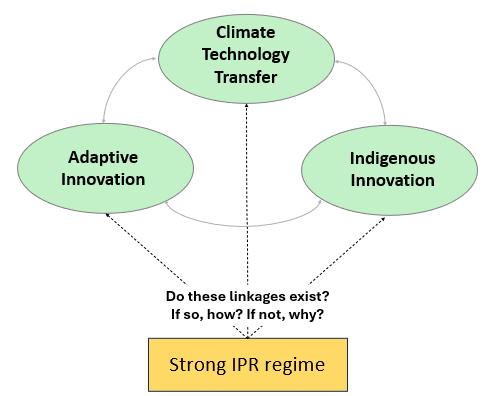}
			\caption{Research questions and conceptual building blocks from the review}
			\label{fig:RQ}
		\end{figure}

		Figure \ref{fig:RQ} depicts our main research questions positioned within the key concepts derived from the literature review. In analytical terms, there is a difference between innovation primarily driven by local knowledge, expertise, and challenges (indigenous innovation) and follow-on innovation involving improvements to imported technologies via ITT (adaptive innovation). In practice, however, there is no clear-cut distinction between both, since indigenous innovation may also build on imported knowledge. Further, there may be a positive feedback loop between indigenous and adaptive innovation: success in one channel may feed into and promote success in the other channels \citep{fu2011role}.
		
		\section{Methodology}
		\label{sec:med}
        Our methodology uses exploratory research \citep{stebbins2001exploratory} to develop new insights on the diffusion of climate technologies amongst developing nations. Exploratory research is appropriate given the little empirical evidence on this topic \citep{maskus2009intellectual, ring2021patent, ipcc2022climate}. We examine the complex relationship between IPRs, climate technology transfer, innovation, and sustainable development in developing countries through in-depth case studies, covering Bangladesh, India, Kenya, and South Africa. Each country serves as a unique instrumental case study, allowing for deeper insights \citep{stake1995art}. 
        
        We conducted 20 semi-structured interviews with key stakeholders, guided by the literature review, identified knowledge gaps, and project objectives. 
        The overall sample size is small, but this is common for exploratory studies where there is little available evidence \citep{stebbins2001exploratory}. While the sample size places limitations on the generalizability of this study’s claims \citep{payne2005generalization}, our case study research aims to produce an intensive examination of cases to inform wider research \citep{stake1995art}. The findings complement our literature review to outline areas for future research in Section \ref{sec: discussion}.\footnote{The interview guide is provided in \ref{app:interview_guide}.} 
		
		\subsection{Case selection}
		\label{case selection}
		Case study sampling techniques select research subjects based on their characteristics, unlike the random sampling methods used in most quantitative research studies \citep{stake1995art}. The selection was done in two steps: (1) We selected countries relevant to our research questions, and (2) we chose individual interviewees from each country following \citet{merriam2015qualitative}.
		
		We selected countries with diverse development levels and economic structures. Using World Bank indicators, we chose Bangladesh, India, and Kenya as lower middle-income nations, with Bangladesh also classified as a Least Developed Country (LDC) by the UN. Additionally, we studied South Africa, an upper-middle-income nation heavily reliant on fossil fuels. We chose pairs of neighbouring countries (India and Bangladesh; South Africa and Kenya) with varied economic structures and development challenges, promoting easier and more meaningful comparisons. The selection of countries with similar legal systems (common law backgrounds in the British Commonwealth tradition) facilitates meaningful comparisons.
		
		We interviewed five experts from each country, including individuals from national IPR offices, government officials, local and foreign climate technology firms, and academic/policy experts in areas like climate innovation, IP law, and sustainable development. Further details of the interview participants are in \ref{app:parti}.
		
		\subsection{Data collection and analysis}
		We contacted interviewees via email or telephone, provided background information, and obtained appropriate consent. Then, we conducted 45-60 minute video interviews, and transcribed them, ensuring privacy. Interviewees were given pseudonyms using randomly generated codes. Thematic analysis was completed using NVivo software \citep{clarke2021thematic}.
		
		The data was initially coded line by line, then grouped under recurring themes using grounded theory methods \citep{strauss1987qualitative, bryman2016social}. Thematic analysis was guided by emerging themes in the data, findings from the literature review, and relevant policy documents on IPRs and sustainable development.
		
		\section{Findings}
		\label{sec:results}
		
		\subsection{Bangladesh}
		Bangladesh is a lower middle-income country, with a GDP per capita of USD 2,457 and a population of 169.4 million (2021). The national agenda emphasises employment and education, driven by the substantial proportion of young people. Agriculture is the dominant sector, employing approximately 37.7\% of the population \citep{climateriskbangladesh}, but the country aims to transition its economy from its current agriculture-based model towards manufacturing and services. Given the expected rise in emissions during this economic restructuring, interviewees highlighted clean energy and transportation as crucial mitigation areas, requiring significant investments \citep{tna2012bangladesh} with opportunities for technological leapfrogging.
		
		Although priority areas for mitigation exist, Bangladesh currently places more emphasis on adaptation \citep{ndc2021bangladesh, tna2012bangladesh}. Its tropical location in a river delta characterised by seasonal monsoons makes the country highly vulnerable to extreme weather events such as floods, heatwaves, and cyclones. Adaptation in agriculture is highly significant, given its substantial contribution to the economy. While Bangladesh prioritises adaptation, it also aims for adaptation solutions with mitigation co-benefits \citep{e}.
		
		Interviewees characterised Bangladesh’s current IPR system as immature, paper-based, underutilised, and difficult to access. A civil servant stated that at \emph{`university level or secondary level, most of our students and even teachers are not aware of intellectual property; how to use [it], how to manage, and how to develop a product within the framework of IP'}. There was also a perception that the public was relatively \emph{`indifferent'} to IPRs, which may be due to lacking awareness and enforcement. One entrepreneur described an experience of visiting a factory where drawings of patented inventions were openly copied.
		
		However, the Bangladeshi IPR system is currently in a period of change, intending to encourage a broader uptake of IPR protection and promote innovation. When the interviews were conducted, new legislation in the form of the Bangladesh Patent Act 2022 had been recently enacted. This repealed the patent provisions of the Patents and Designs Act, 1911 and made separate provisions for patent protection. The 2022 Act incorporated substantive examination of patent applications and extended protection from 16 years to 20 years.  Moreover, the 2022 Act introduced a utility model (\emph{`petty patent'}) to accommodate \emph{`low-tech'} solutions that would not meet the standard for patents.\footnote{Bangladesh Patent Act 2002, s 32.} Like patents, utility models provide inventors with the exclusive right to prevent others from commercially exploiting their inventions, but with a shorter term of protection than patents, typically six to ten years. 
		Interviewees mentioned that the utility model system has the potential to stimulate indigenous innovation in locally-specific adaptation solutions in various sectors, including agriculture, as well as adaptive innovation based on imported mitigation technologies in the transport and energy sectors to improve and adapt existing solutions.  
		
		Since the interviews were conducted, the 2022 Act has been replaced by the Patents Act 2023. The 2023 Act has (judged by Western standards) some unconventional features and appears to be designed with the interests of the Bangladeshi pharmaceutical sector very much in mind. It retains the new features outlined above, but raises the threshold for obtaining utility models and reduces the term of protection to 8 years (from the 10 years set out in the 2022 legislation).\footnote{Bangladesh Patents Act 2023, s 42. We thank Ataul Karim for providing a translation and analysis of this provision.}  
		
		Interviewees agreed that IPRs can give foreign investors a sense of security in their investment, as argued in the literature (see section \ref{subsec:IPR and ctt}). However, it was also said that final investment decisions would be driven by the business case and market conditions. One foreign entrepreneur in Bangladesh highlighted that \emph{`weak IPR enforcement in Bangladesh is unlikely to discourage technology transfer if there is a strong enough business case to enter the market, and it seems that foreign companies are aware of and willing to accept the risks associated with entering the market in a low-income country'}. Other barriers to ITT that were raised included a lack of local expertise in implementing or maintaining climate technologies and a lack of knowledge about solutions available on the international market that could be applied in Bangladesh.
		
		Despite the overall low level of awareness and general scepticism about the current role of IPRs in Bangladesh, respondents indicated that trademarks are relatively widely used because the system is user-friendly and appropriate for the economic setting. Trademarks are relevant to many small- to medium-sized enterprises as a means of establishing a brand. In addition, regulatory requirements relating to standards require traders to register their trademarks with the Bangladesh Standards and Testing Institution. Securing a trademark registration was considered to be much easier than applying for a patent.
		
		\subsection{India}\label{R_India}
		India is an example of a lower middle-income country with a GDP per capita of USD 2,256 and a population of 1.39 billion (2021). Global efforts to meet the 1.5-degree target depend on India successfully tackling its emissions, given its projected population and economic growth over the coming decades. India’s priority areas for mitigation are energy, transportation, and waste management. The key area for adaptation is agriculture, where farmers face drought and excess water due to longer and less predictable monsoons. 
		
		Despite India’s reliance on fossil fuels, interviewees were optimistic about India’s ability to transition to clean energy, referring to the rapid deployment of solar power and one of the world’s largest renewable expansion programmes (especially solar, hydro, and onshore wind). India is expanding its support to accelerate the adoption and production of hybrid and electric vehicles, including e-rickshaws, which account for about 60\% of EV sales in the country. To accelerate mitigation effects and to facilitate climate technology transfer, India’s Nationally Determined Contributions explicitly assert that developed countries should actively engage in R\&D collaboration with India and exempt developing countries from IPR costs \citep{ndc2022india}. India advocates for its \emph{`right to grow'}, often being thought of as using the status as a developing country to avoid stringent climate commitments. 
		
		Aligned with the emphasis on economic growth, India has undergone an ideological shift from being conventionally antithetical to IPRs to a more nuanced approach that accepts that strong IPRs may have a role in fostering industrialisation. The shift was particularly pronounced in the pharmaceutical and agricultural sectors. Since the early 2000s, the Indian pharmaceutical sector was said to have moved from the manufacturing of generic, off-patent drugs to an innovative and competitive industry that patents its own inventions. Moreover, India was initially opposed to granting IPRs for GMOs, but it recently adopted a more permissive stance, approving an increasing number of patents for GMO inventions (e.g. GMO mustard seeds).
		
		Given these conditions, interviewees mentioned that the IPR system in India is relatively well-established given its stage of development. It was said that \emph{`India is the most successful at having intellectual property policies that suit their national development objectives'}, referring to the previous Prime Minister, Manmohan Singh’s backing of provisions for compulsory licensing under the TRIPS Agreement in the lead up to the Doha Declaration on public health in 2001.\footnote{Agreement on Trade-Related Aspects of Intellectual Property Rights, Apr. 15, 1994, Marrakesh Agreements Establishing the World Trade Organization, Annex 1C, 1869 U.N.T.S. 299 (TRIPS Agreement).}
		
		However, despite being relatively well-established, on-the-ground experiences from interviewees suggest that India's IPR system remains challenging to use due to a cumbersome bureaucracy, weak enforcement, and low public awareness. Similar to Bangladesh, interviewees in India viewed trademarks as more user-friendly and suitable for promoting technology adoption and entrepreneurship than patents, which were thought to be overly technical and obscure. 
		
		One entrepreneur, who developed a rainwater management system for rural female farmers, used a trademark to establish a brand for this system. This was said to have helped encourage adoption in poor communities with low levels of literacy. 
		One interviewee stated that they did not want to prevent copying if allowing infringement would foster climate technology uptake, describing Gandhian innovation as driven by impact rather than profit: \emph{`serving the last person in the queue in the best possible way. My objective of this innovation, rather than making money or rather than earning money through patent rights, I was more eager to make the impact maximum.'}
		
		As for the technology transfer from foreign investors, IPR protection was identified as having some relevance in encouraging FDI in India, as many investors at least look to IPR protection as a \emph{`tick in the box'} when making their decisions. However, it was noted that other legal barriers impede the adoption of foreign technology in India. For instance, one of the interviewees (a foreign entrepreneur who set up a business presence in India) described it being difficult to navigate the complex legal rules to set up a company in India as an outsider without local contacts or knowledge about local rules and regulations. 
		Businesses need an on-the-ground presence, such as large-scale domestic investment or working with local partners, because it gives them a network to manage bureaucratic hurdles in India. 
		
		Interviewees believed that the current IPR system is inadequate to encourage indigenous or adaptive climate technology development in India. This perception arises because many of the required climate innovations in India so far have been low-cost and incremental, which struggle to meet the novelty and non-obviousness requirements for patent protection. How IPRs incentivise expensive solutions was seen as conflicting with local needs, particularly regarding solutions relevant to agriculture or poorer communities in rural areas. Moreover, some interviewees expressed the attitude that discussions about IPRs may only be relevant after the demand for mitigation technologies had increased through more stringent environmental regulations.
		
		In sum, the relevance of IPRs for climate technology transfer and innovation in India indicates a substantial sectoral difference in the coming decades. While remaining irrelevant in many areas, for some technologies such as GMOs for agricultural adaptation or pharmaceutical inventions for health adaptation, the IPR system seems increasingly relevant. 
		
		\subsection{Kenya}
		Kenya is a lower middle-income country, with USD 2,006 GDP per capita and a population of 54 million (2021). Kenya’s key sectors are agriculture and tourism, which are highly vulnerable to climate risks. The Natural Disaster Statistics from the World Bank \citep{WB2021kenya} show that drought is the main risk area threatening water and food security in the country, followed by riverine floods and epidemics. Relevant indigenous innovations to address these issues were identified as critical, including medicines or crops that can cope with Kenya’s specific environment. Given the importance of locally-specific indigenous innovation, technology transfer between countries in East Africa with similar climate risks was mentioned as taking place informally, with efforts being made to create formalised commitments to identify synergies to support technological capacity. 
		
		Interviewees mentioned that the established system of utility models in Kenya has been actively used with the expectation that it will encourage high-impact, low-cost and locally-specific innovation, resulting in many more utility model applications than patents. The COVID-19 pandemic was described as an event that triggered widespread domestic innovation to provide low-cost solutions, with parallels to climate change. Kenyan interviewees noted that there was also greater use of trademarks domestically, particularly by \emph{`the big players, not the small players'}, to mitigate concerns about the quality of imported products. 
		
		Despite the relatively widespread use of utility models and trademarks, participants noted that the IP system in Kenya, particularly the patent system, is still young and that greater public awareness is needed. IP was described as \emph{`legalistic'}, \emph{`inaccessible'}, \emph{`bureaucratic'} and \emph{`paper-based'}. Businesses seeking patent protection find the system lengthy, complex and not user-friendly. One entrepreneur commented that \emph{`the process is like a mirage, it's really not clear'}. It was also reported that many universities do not have a policy on IPR arrangements concerning their research. Unsurprisingly, when asked about the conditions for copying and enforcement, the environment was described as a \emph{`wild west'}, with widespread infringement of IPRs, particularly trademarks, which are widely used in the country. This lack of IPR enforcement may be due to the stage of development. One participant suggested that \emph{`Kenya has other problems to deal with before it deals with IPRs'}.
		
		While Kenya’s priority agenda is adaptation, there are mitigation areas where follow-on adaptive innovation is needed, building on imported solutions. Kenya’s progress in renewable energy was seen as encouraging, so far, mostly based on hydropower. However, interviewees reported that hydropower is coming under pressure due to increasing droughts, leading to water shortages affecting agriculture. As a result, Kenya is turning to solar energy and biogas, given the abundance of sunlight and biogas on farms. The country’s potential for leapfrogging was discussed in the areas of off-grid energy and clean transport. In particular, the development of cheaper e-mobility vehicles for urban areas was expected to become competitive with traditional fossil fuel vehicles, as e-vehicles can be manufactured domestically at lower costs, whereas most traditional vehicles face importation costs. 
		
		IPRs were not at the forefront of the drivers of FDI or the decision to enter the Kenyan market, given the country’s current technological capabilities. Participants stated that foreign firms face lower infringement risks when bringing climate technologies to Kenya because of its lower capacity to reverse-engineer imported technologies, especially in the case of heterogeneous and complex renewable energy technologies. Conversely, this lack of technical expertise in Kenya poses a challenge to adopting climate technologies that rely on local skills. Interviewees were critical of climate technology transfer from developed countries to Kenya, arguing that prevailing international initiatives for ITT are inadequate and ineffective.
		
		\subsection{South Africa}
		South Africa is an example of an upper-middle income country, with a GDP per capita of USD 6,994, which is very unequally distributed, and a population of 59.3 million (2021). The country’s energy and economy heavily rely on coal-based infrastructure, with only 10.5\% of energy consumption based on renewables in 2019. Given this, the country’s priority climate agenda is the energy transition by 2030, followed by the transition in the transport and hard-to-mitigate sectors by 2040. South Africa has been described as facing a double burden of meeting high expectations for its response to the climate emergency as the second-largest economy in Africa with high per-capita emissions, while still being a \emph{poor country}, without sufficient resources to do so at the pace expected by the global community. 
		
		Interviewees pointed out local knowledge of available technologies is insufficient to determine which mitigation solutions best suit South Africa’s needs. Discussions were dominated by the prevailing blackouts and energy crisis, affecting the country’s economy as a whole, particularly the manufacturing and agricultural sectors that rely on a stable electricity supply. These instabilities in the power grid are, to some extent, encouraging the uptake of electric battery systems.
		
		South Africa is more focused on mitigation than other developing countries, but it also has significant adaptation needs. The main risks are unexpected floods and droughts, with subsequent sector-wide water scarcity and sanitation problems in health, agriculture, and settlements. The economy heavily relies on industrial agriculture, which uses about 50\% of scarce water resources.\footnote{The Natural Resources Defense Council in the US defines industrial agriculture as \emph{`the large-scale, intensive production of crops and livestock, often involving chemical fertilizers for crops or the routine, harmful use of antibiotics for livestock'}.} 
		Adaptation in agriculture is therefore urgently needed. These include hazard early warning systems, climate-smart agriculture and new crop varieties that are less vulnerable to changing climate conditions. Many of these adaptation solutions are low-complexity indigenous innovations based on local knowledge. In contrast, other adaptation technologies, such as climate-resilient crops or new therapeutics, typically require substantial R\&D investment and are largely based on existing solutions transferred from developed countries.
		
		Respondents indicated that the IPR system in South Africa is relatively well established, but that the awareness of IPRs among the public is low and IPRs are perceived as \emph{`legalistic'} rights for the elite. There appeared to be quite active IP training and capacity-building programmes run by the South African Intellectual Property Office to promote awareness and understanding of IPRs. Interestingly, patents in South Africa are granted automatically if they meet procedural requirements without substantive examination.\footnote{\emph{`Substantive examination'} in this context refers to a system in which patent examiners working within the patent office assess whether the claimed invention clears the legal thresholds for patentability, in particular, through assessment of whether the invention is novel, involves an inventive step (is non-obvious), is capable of industrial application, and is sufficiently disclosed.} 
		
		On the one hand, this automatic patent system can encourage patenting by minimising bureaucratic hurdles. On the other hand, without a formal examination process, applicants and potential investors may have little confidence in the validity of a patent and whether it can be enforced. 
		As this threatens to undermine foreign investor confidence, one interviewee noted that the South African Intellectual Property Office is preparing to introduce substantive examination in the coming years. 
		
		Interviewees reported significant interest from foreign investors in the South African market and unprecedented interest in renewable energy projects on the African continent. One noted \emph{`a lot of investment coming in from a lot of different angles, it’s definitely something I think every single person at my company can agree is something we’ve never seen before'}. Interviewees generally agreed that IPRs are important for foreign investor confidence. In particular, IPR enforcement in South Africa was reported to be particularly important for trademarks as a means of tackling counterfeiting. 
		
		Although participants agreed that weak IPR enforcement could harm foreign investor confidence, there was some resistance to the idea that FDI is IPR-driven. One interviewee noted that there was little evidence regarding the impact of IPRs on technology transfer to South Africa, so one could only \emph{`speculate'} on their possible benefits at this stage. In addition, another interviewee noted that the sectors in which foreign investors are currently most interested, such as mining, are often far from sophisticated technologies protected by patents.
		
		Respondents felt that domestic innovation in South Africa was not strongly influenced by IP protection, either positively or negatively. One interviewee noted that \emph{`IPR had little impact on local innovation before South Africa signed the TRIPS Agreement or afterwards'}. It was felt that in a country with a significant gap between rich and poor, prevailing knowledge about IPRs remains concentrated among the elite. Moreover, for some renewables, such as solar and wind, adoption and scale-up are the main issues, as most of the relevant key patents have already expired.
		
		Participants also noted that the establishment of IPR protection in South Africa was seen as promoting the adoption of IPR systems on the African continent, referring to South Africa’s position in the negotiation of the IP chapter of the African Continental Free Trade Agreement. For sophisticated and early-stage climate solutions (e.g. hard-to- mitigate sectors such as clean manufacturing), there are potential areas where IP may be relevant to mitigation in the future.
		
		\FloatBarrier
		\section{Discussion}
		\label{sec: discussion}
\begin{figure}
			\centering
			\includegraphics[width=1.0\textwidth]{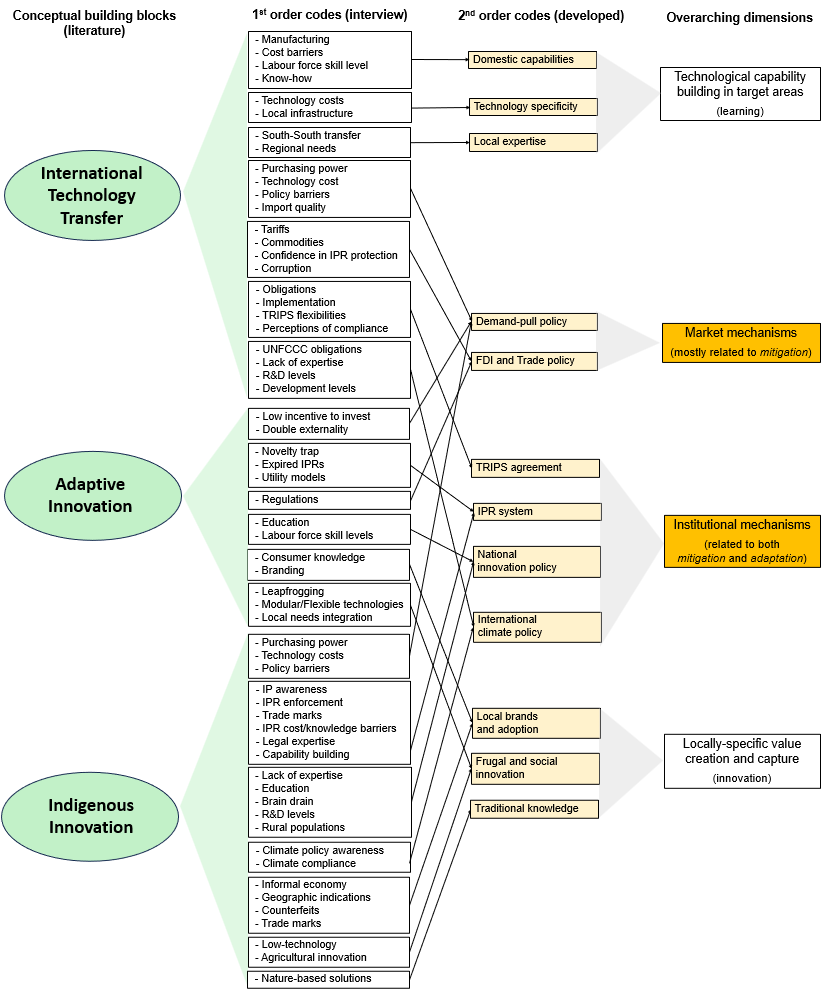}
			\caption{Process of coding and thematic analysis}
			\label{fig:theme}
		\end{figure}

		From the interviews, it emerges that IPRs currently play an insignificant role in climate technology transfer and innovation in developing countries. However, there is scope to make IPRs work for climate and sustainable development goals. We reconcile and contextualise the insights from the literature and interviews, and derive policy implications related to IPRs and beyond, structured along the three domains: international technology transfer (Section \ref{subsec:results discussion transfer}), indigenous innovation (Section \ref{subsec:results discussion indigenous}), and adaptive innovation (Section \ref{subsec:results discussion adaptive}). Figure \ref{fig:theme} summarizes the process of thematic analysis, starting from the literature review to semi-structured interviews and subsequent analysis. 

		\subsection{International technology transfer} 
		\label{subsec:results discussion transfer}
		ITT involves market and institutional channels. Market-based ITT is a by-product of FDI and trade. Many policy discussion on climate-related ITT are also related institutional support mechanisms within the UNFCCC. Below, we discuss the role of the different ITT channels by area of national and international policymaking, focusing on IPR-related aspects.  
		
		
		\subsubsection{Market mechanisms}
		\label{subsec: discussion market}
		\paragraph{FDI policy}
		Interviewees acknowledged the role of IPRs in FDI, especially in knowledge-intensive sectors. Yet, they considered them minor in FDI decisions, a formality, a \emph{`tick in the box as part of the investment decision'}. Market opportunities were far more influential. 
		
		However, FDI projects were criticized for lacking knowledge exchange, focusing primarily on market exploration and resource exploitation: \emph{`it’s the company [...] coming and establishing itself, and it’s not really a transfer of technology'}. Most countries lack absorptive capacities for knowledge-intensive FDIs and associated ITT \citep[][and see also section \ref{subsec:results discussion adaptive}]{mathew2023role, olawuyi2018technology, li2011sources}. Making FDI an effective route of ITT requires co-production and co-development efforts to assist developing countries in acquiring technological capabilities \citep{lee2015comparative}. Effective IPR systems become crucial for ITT via FDI once domestic technological capacity reaches a certain level.
		
		Our results align with previous studies showing that market size, costs, infrastructure, and growth are more important than IPRs (Section \ref{subsec:IPR and ctt}). China and India attracted significant FDI despite weak IPR enforcement \citep{awokuse2010intellectual, rai2008effect, tikku1998indian}. There may, however, be a positive relationship between robust IPR protection and FDI in technology-intensive sectors, where FDI goes beyond low-cost manufacturing or primary resource extraction and includes elements of joint ventures and R\&D collaboration \citep{abdel2015intellectual, leahy2010intellectual}. 
		
		Many low-carbon projects risk continuing the trend of resource-focused FDI. Wind, sun, land and certain raw materials are the primary resources of many mitigation technologies. Policy can stimulate knowledge flows in FDI, for example through training obligations, local content requirements, and joint ventures \citep{maskus2010differentiated, corvaglia2014south}. Yet, the success of such measures depends on realistic targets and a broader industrial policy \citep{scheifele2022impact, fu2011role, cheung2004spillover, wong2006technovation}.
		
		\paragraph{Trade and demand-pull policy} 
		Interviewees noted \emph{`at the end of the day, the key constraint for developing countries is cost'}, especially labour and production costs, when sourcing from developed countries, for both IP-protected and off-patent climate technologies. Climate technologies from China and India are becoming preferred alternatives due to their affordability \citep{quitzow2015dynamics, lema2012technology}. 
  
        Although IPRs theoretically increase the cost of climate technologies, many mitigation technologies are traded in highly competitive markets and substitute technologies without patent protection are available. Nevertheless, demand for climate technologies remains lower than desired, due to cost barriers. Past studies highlighting IPR as partly causative of these barriers may be outdated \citep{suzuki2015identifying}. 
		
		To overcome cost-related obstacles, tariff reductions in Preferential Trade Agreements (PTAs) for relevant technologies could incentivize adoption \citep{de2020barriers, brandi2020environmental, corvaglia2014south, brewer2008climate, maskus1998role}. However, the definition of eligible climate-related goods in PTAs needs careful consideration to avoid the effective exclusion of products from developing countries \citep{de2020barriers}.
		
		Additionally, promoting trade \emph{`among'} developing countries, focusing on technologies suited to local contexts, could be a viable solution. Examples include the e-rikshaw and off-grid renewables. Certain agricultural technologies have the greatest benefits in neighbouring countries with similar climates or agricultural production systems. As one interviewee said, \emph{`what works in Bangladesh can also work in Sri Lanka'}. To date, international ITT policies tended to focus on developed countries as exporters and are often ill-suited to promote ITT between less developed countries. Trade among developing countries is often informal, and initiatives to coordinate these practices are lacking. There is little empirical record to understand their scale and efficacy \citep{bouet2018informal}. 
		
		Patent protection is often lacking for products developed in developing countries. This is connected to concerns about developing countries becoming `dumping sites’ for low-cost, low-quality climate solutions exported from China. Against this background, \textit{trademarks} and \textit{branding} are seen as quality assurance markers that are accessible to local entrepreneurs looking to protect their adaptive and indigenous inventions (Section \ref{subsec:results discussion indigenous}-\ref{subsec:results discussion adaptive}).
		
		\subsubsection{Institutional mechanisms}
		\label{subsec: discussion institutions}
		International climate and trade agreements both incorporate institutional ITT provisions, but their effectiveness is debated \citep{uddin2020international}. Interviewees described specific flaws in the underlying legal frameworks.
		
		\paragraph{TRIPS Agreement}
		The main flaw in current ITT provisions is their disregard for developing countries’ \textit{absorptive capacity}. While Article 66.2 and Article 67 of the TRIPS Agreement obligate developed countries to transfer technology and incorporate IPR into local innovation systems of developing nations, these measures are often ineffective due to limited local capacity. Quoting one interviewee: \emph{`they give us some technology and some training, but not the technology transfer in the right way'}. 
		
		\paragraph{International climate policy}
		To address the limited capacity issue, the UNFCCC introduced the Joint Implementation and the Clean Development Mechanism (CDM) initiatives, which are designed to support climate-related FDI and joint ventures. However, these performed poorly in ITT and emission reductions and were highly concentrated in four countries (China, India, South Africa, and Brazil) with low participation rates of most developing countries \citep{kainou2022collapse, lo2022emission, corvaglia2014south}. Integrating local partners through well-calibrated local content and training requirements and providing incentives could enhance their effectiveness.
		
		\paragraph{Harmonising WTO and UNFCCC}
		Another flaw lies in the division of responsibilities between the WTO and UNFCCC into separate international policy domains \citep{cheng2022intellectual, zhou2019can}. Participants noted that both the UNFCCC and the TRIPS Agreement impose obligations on technology transfer, but it is less clear how these provisions interact. 
  
        As one interviewee stated, there is a need for greater \emph{‘synergy between intellectual property and the issues and challenges of the WTO and the international agreements related to climate change’}. Another interviewee suggested it would be difficult to tackle this problem due to \emph{`lack of hierarchy between the [respective] treaties'} in international law.  Uniform interpretation would be difficult to obtain given the \emph{`proliferation of plurilateral arrangements and agreements'}, and the rise of \emph{`alliance multilateralism'}, or regional trade agreements, amongst powerful trade blocks. 
		
		In principle, the two bodies could work together more closely. For example, the UNFCCC could help guide the interpretation of TRIPS provisions for climate technologies, potentially by aligning them with its Technology Needs Assessment \citep{de2015technology}. 
  
        However, beyond the inevitable barriers to cooperation created by institutional territoriality, it should also be noted that the UNFCCC and WTO are rooted in different economic traditions. The UNFCCC is motivated by the institutional tradition, embodying the principle of CBDR. In contrast, the WTO is shaped by the free trade tradition inspired by economic liberalism. Although these different theoretical foundations may further complicate institutional cooperation, an alternative perspective would be to see them as playing different roles in promoting the transfer and innovation of mitigation (WTO) and adaptation technologies (UNFCCC) (see Figure \ref{fig:discussion_overview}).
		
		\paragraph{Technological capacity building}
		Existing institutional mechanisms have a limitation: their focus on ITT from developed to developing countries neglects technologies originating from developing nations \citep{corvaglia2014south, brewer2008climate}. While this approach aligns with the UNFCCC’s principle of CBDR, it ignores the potential of South-South ITT. This topic gains increasing traction in the literature \citep{uddin2020international, urban2018china, corvaglia2014south, brewer2008climate}, and the UN has now incorporated it into its agenda.\footnote{\url{https://news.un.org/en/story/2019/03/1035011} [Last accessed 29 August 2023].} 
		
		Support from developed countries may, however, still be pivotal in facilitating ITT among developing countries. \citet{uddin2020international} emphasizes the significance of triangular cooperation. This is where developed countries can assist \emph{`Southern-driven partnerships between two or more developing countries supported by a developed country(ies) or multilateral organization(s) to implement development cooperation programmes and projects'} (UN, 2012).\footnote{The UN defines South-South cooperation as a \textit{`process whereby two or more developing countries pursue their individual and/or shared national capacity development objectives through exchanges of knowledge, skills, resources and technical know-how, and through regional and interregional collective actions, including partnerships involving governments, regional organizations, civil society, academia and the private sector, for their individual and/or mutual benefit within and across regions'} \citep[see][]{UN2012framework}.} This support can be financial, legal, or technical. 
		
		Infrastructure assistance is important in this context. Many developing countries lack existing fossil fuel infrastructure, providing an opportunity to adopt renewables without transitioning away from carbon-intensive sources. Financial support for relevant infrastructure can catalyse the widespread adoption of locally adapted or invented clean solutions. This potential aligns with opportunities for leapfrogging, illustrated by the success story of rapid mobile phone technology adoption \citep{ahmad2020mobile, aker2010mobile}. 
		\FloatBarrier
		
		\subsection{Indigenous innovation}
		\label{subsec:results discussion indigenous}
		\paragraph{Locally-specific value creation and capture}
		Many climate solutions required by developing countries diverge from those needed in advanced economies. Local entrepreneurs will therefore need to devise effective solutions tailored to local challenges, including low-tech and NbS.
		
		Patent law has little applicability to NbS. Consider, for example, projects involving the restoration of mangroves, which can sequester carbon dioxide, preserve biodiversity and protect coastal areas from adverse weather events \citep{arkema2023evidence}. Such projects can be impactful and do not implicate patent law. As regards low-tech solutions, patent law’s strict novelty and non-obviousness standards mean that patents can play no role in incentivising the development of simple but impactful solutions. 
		
		As one interviewee stated, \emph{`the IP system can provide some protection for innovators at the expense of things which may be easier, cheaper, whatever may be more adapted to the Global South'}. Interviewees emphasized the limitations of patents, particularly in the context of \emph{‘frugal innovation’} -- innovative solutions developed under resource constraints \citep{radjou2012frugal, dreyfuss2021technological}. One specific example mentioned by interviewees was oral saline technology for diarrhoea treatment, a high-impact solution that might not meet the stringent novelty requirements of patents but can make a significant difference in developing countries.
		
		
		\paragraph{National innovation policy and IPR system}
		
		Interviewees from Bangladesh and Kenya pointed out that IPRs are largely owned by foreign firms, consistent with the literature \citep{corvaglia2014south, united2010patents}. An interviewee noted, \emph{`intellectual property comes after technological development'}, emphasizing that IPR protection becomes a relevant part of well-functioning innovation systems after their establishment. Moreover, interviewees from India, with a more established IPR system, attribute India’s innovation to demand-side incentives rather than IPR protection, aligning with a quote from Kenya \emph{`innovation has always been there, it’s the law that comes now'}.
		
		Trademarks, however, are widely used by local innovators and more equally distributed among foreign and domestic applicants than patents \citep{suthersanen2006utility, baroncelli2005global}. Compared to patents, trademarks are easier to register, which may promote low-cost climate innovation and local entrepreneurship. They can help build consumer loyalty and trust, encouraging the creation of reliable and functional climate solutions. 
        The relative reliance on trademarks compared to other IPRs correlates with development \citep[see also][]{kang2020intellectual}. One respondent commented that in all economies the adoption of IPR systems follows a common trajectory as the economy develops: \emph{`you start with trademarks, inevitably, because trademarks have existed since markets existed...slightly later when markets start to develop...you look at utility models and patents'}.

		\FloatBarrier
		\subsection{Adaptive innovation}
		\label{subsec:results discussion adaptive}
		
		
		\paragraph{Locally-specific value creation and capture}
		Adaptive innovation can beneficial for local society and help speed up diffusion of existing technology. For example, the rise of e-scooters, e-rickshaws, and off-grid solar PV solutions in various developing countries show the great potential for locally-specific markets for adaptive innovation building on imported technology \citep{urmee2016social, akamanzi2016silicon, lema2012technology}. Adaptive innovation also offers opportunities for sustainable growth and leapfrogging by skipping the emission-intensive phase of industrialisation. 
		
		\paragraph{National innovation policy and IPR system}
		Adaptive innovation is the channel of climate technology diffusion where IPRs matter (both positive and negative). On the one hand, follow-on innovation that involves significantly adapting or repurposing cutting-edge imported technologies can clear the novelty trap. This increases the potential for patents to incentivise local innovators including, potentially, by guaranteeing returns in export markets in which adaptive innovations are adopted. On the other hand, adaptive innovations may be hindered if they risk infringing IPRs over imported technologies, namely, in situations where patented imported technologies are being copied, changed and repurposed. 
		
		Our results suggest that adaptive innovation is not yet sufficiently widespread in any of the countries studied for IPRs to be a relevant consideration. This should not be a matter of surprise. 
        National policy for adaptive innovation needs to focus on local capability formation, which can form a basis for technology leap-frogging. IPRs are, as yet, something that is of relatively little concern, but this may change. As one interviewee stated, \emph{`[w]e have many high-tech solutions in rich countries but what’s needed are low-tech solutions, but if there is a barrier to protect these inventions and if [IPR] isn’t a barrier to local development, then it might be a barrier to upscaling green technology diffusion'}. But equally, it is possible that IPRs play a more positive role in adaptive innovation. At present, it would be premature to politically intervene in either direction. 

        As an alternative, mentioned by various interviewees and in agreement with existing studies, \textit{utility models} can offer an incentive for the development of low-cost and low-complexity climate technologies \citep{dreyfuss2021technological, kim2012appropriate, suthersanen2006utility}. However, the efficacy of second-tier systems for protecting innovation is controversial. Some countries that have experimented with such systems have since abandoned them. For example, in Australia, the \emph{innovation patent} started being phased out in August 2021.\footnote{See Intellectual Property Laws Amendment (Productivity Commission Response Part 2 and Other Measures) Act 2020 (Cth).} Whether this experience holds lessons for developing countries is to be determined, as utility models might play a positive role in catching-up \citep{suthersanen2006utility, kim2012appropriate}. 
		
		\FloatBarrier
		\section{IPRs as part of a broader strategy for climate technology in developing countries}
		\label{sec:discussion mitigation adaptation}
		IPRs are not going to save the world, nor are they going to doom us to disaster. Getting IPR policy right, however, can make a difference at the margins of tackling climate change. Current debates have been dominated by whether IPRs should be weakened to facilitate ITT -- especially by expanding TRIPS waivers to climate technologies. Our analysis suggests that this debate misses the mark. Weakening IPRs is unlikely to facilitate ITT to developing countries, and might even do harm if it were to shake investor confidence. The bigger danger, however, is that the debate about a TRIPS waiver will become a distraction. As one interviewee noted, the concern is that this debate is a \emph{`storm in a teacup'} that \emph{`misdirected critical energies towards the wrong targets'}
		
		Our interviewees suggested that domestic demand-pull policies should be prioritised to strengthen the demand for ITT. This conclusion is in line with emerging discussions on the importance of demand-focus rather than supply-focus in policies supporting sustainability transition \citep[e.g.][]{boon2018demand}. Demand-focused policies in this context include environmental regulation and financial incentives, with the latter including subsidies, public procurement programmes and carbon pricing, but recognizing that scarce public budgets remain a barrier to fiscal support in poor countries. 
		
		Local demand is key and a precondition to attract FDI, while IPR protection has little relevance to ITT in mitigation. However, robust national IP policies may provide some confidence to foreign investors, which can facilitate knowledge transfer and capacity building. Our interviewees emphasised the need for FDI to be accompanied by realistic and specific plans for the actual transfer of skills and capabilities. The local IPR system may have a limited but positive role to play alongside, for example, mandatory joint venture, local sourcing and collaboration requirements.
		
		Turning to indigenous and adaptive innovation, the focus should be on local capability formation. Building technical expertise and ensuring that finance is available to local entrepreneurs should be prioritized. It is, therefore, unsurprising that one interviewee described weakening IPRs, rather than funding capacity-building or taking similar measures, as a \emph{`cynical minimalist approach'}. Meaningful measures are more likely to come in the form of funding to implement simple but impactful solutions, and formalised systems for targeted knowledge transfer. Once local businesses can produce indigenous and adaptive innovations, the creation and expansion of domestic and international markets for these technologies become important. 
		
		Interviewees agreed that trademarks already play an important role in facilitating domestic market penetration. Indeed, it is striking that interviewees from all four countries highlighted the role of trademarks in helping to create brand loyalty. There is no obvious reason why trademarks could not perform the same role in cross-border trade if the trend towards South-South technology transfer continues and can be fostered and promoted. Also striking is that several interviewees spoke in positive terms about the potential for utility models to incentivize local innovators and entrepreneurs. Given the somewhat chequered history of second-tier patent systems, this was unexpected, and it is an area where more work is needed.  
		
		None of the above is to suggest that it is not possible to envisage IPRs causing problems for the diffusion of climate technologies in the future. There may be cases in which IPRs slow the diffusion of specific complex mitigation solutions or come to be a barrier to particular forms of adaptive innovation. If such problems emerge, it may be necessary to explore the encouragement of patent pools or compulsory licensing or, indeed, a targeted TRIPS waiver. At present, however, there is no reason to believe that proceeding in any of these directions is necessary or desirable. 
		
		Our study is built around three key channels of climate technology diffusion: (1) ITT, (2) indigenous innovation, and (3) follow-on adaptive innovation. These three routes mutually reinforce, and their relative power depends on the level of technology-specific capabilities. Once essential capabilities have been acquired through ITT, indigenous and adaptive innovation can be fostered through implementing appropriate mechanisms, including a positive role for trademarks and, potentially, utility models. Enhanced domestic capabilities, grounded in indigenous and adaptive innovation, can in turn facilitate the ITT of more advanced solutions. 
		
\begin{figure}
			\centering
			\includegraphics[width=\textwidth]{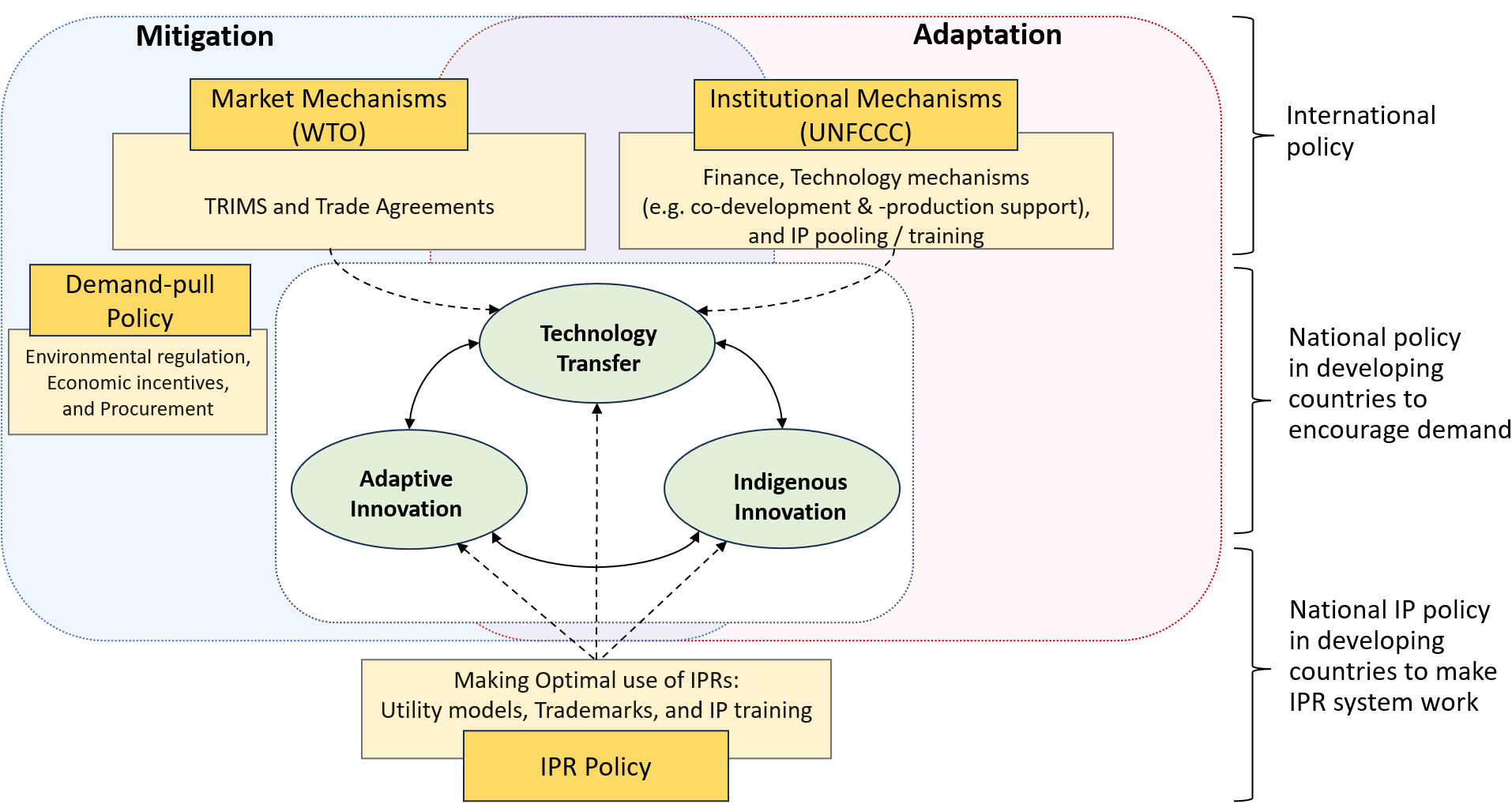}
			\caption{Results in a nutshell}
			\label{fig:discussion_overview}
		\end{figure}

		Figure \ref{fig:discussion_overview} illustrates our understanding of the current landscape, and the interdependencies between different levels of policy, the three areas of climate technology (ITT, indigenous and adaptive innovation), and their relative relevance for mitigation and adaptation in developing countries.

		\FloatBarrier
		\section{Conclusions}
		\label{sec:conclusion}
		The need for climate technologies in developing countries to address existential threats and promote sustainable development has sparked debates about whether IPRs hinder or foster access and use of the required technologies. This study seeks to add to this debate by combining a systematic literature review with an exploratory empirical analysis of 20 in-depth expert interviews in four developing countries. 
		
		Despite our contributions, several limitations persist. First, while we differentiate between mitigation and adaptation technologies based on their intended purpose, it is important to recognise the substantial technological heterogeneity within each category. Policymakers and future studies should therefore consider not only the distinction between mitigation and adaptation, but also the specific technological circumstances, including maturity level, distribution of complementary resources, technological complexity, and IP relevance. 
  
  Second, while this study examined diffusion, future studies may examine pre-market innovation, such as indigenous basic research in developing countries, or local absorptive capacities, such as manufacturing capabilities and business model innovation. 
  
  Third, while significant, our sample size is small, and its findings have limited generalisability. This small sample excludes some perspectives. Future research may broaden the sample of countries or adopt in-depth national case studies. 
  
  Fourth, our findings highlight the importance of trademarks and utility models in lower-income nations. Across the countries studied, trademarks proffer a low-cost brand innovation with possible adoption benefits. Similarly, participants perceived utility models as low-cost alternatives to patents. However, it is an open question whether these schemes will succeed or mirror shortcomings of utility models in technologically advanced economies.

		\newpage
		\printbibliography
		
		\newpage
		\FloatBarrier
		\appendix
		\renewcommand{\appendixname}{Appendix}
		\renewcommand{\thesection}{Appendix \Alph{section}} \setcounter{section}{0}
		\renewcommand{\thefigure}{\Alph{section}.\arabic{figure}} \setcounter{figure}{0}

		\newpage
		
		\section{Interview participants}
		\label{app:parti}
		\begin{figure*}[htp]
			\centering
			\includegraphics[width=10cm, height=10cm,keepaspectratio]{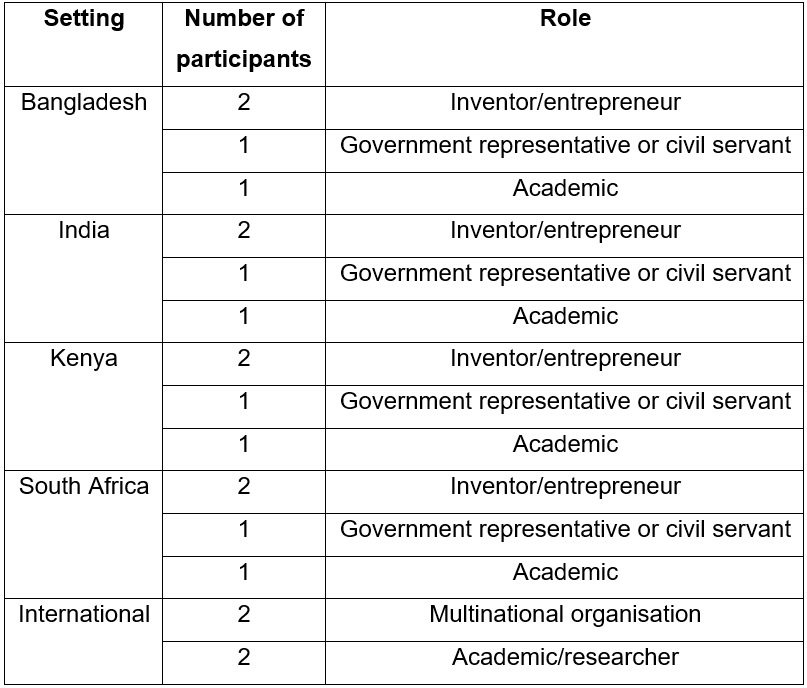}
			\begin{minipage}{\textwidth}
			\end{minipage}
		\end{figure*}
		
		\newpage
		\section{Interview guide}
		\label{app:interview_guide}
		1. To start, could you please outline your expertise, and how you have come across intellectual property law and climate-friendly innovation in your work, if at all?\\
		2. What is your definition of sustainable development?\\
		3. What are the main challenges for sustainable development in your country?\\
		4. What was the impact of the Paris Agreement on sustainable development in your country?\\
		5. How does innovation feature in those challenges? What kinds of innovation or technologies are important for [X country]?\\
		6. Would you say that IPRs are important for domestic innovation? Why or why not?
		\begin{itemize}
			\item What sort of IPRs matter?
			\item Are IPRs important for any other reasons?
			\item *In what ways does IP law impact your business/export decisions? Could you please give an example?
		\end{itemize}
		7. Do you think IPRs are important for international trade? Why or why not?
		\begin{itemize}
			\item *How do IPRs affect foreign direct investment? Would you say it has a positive or negative impact? Why or why not?
		\end{itemize}
		8. Do IPRs facilitate technology transfer?
		\begin{itemize}
			\item From developed economies to the developing world? Between developing nations?
			\item Is technology transfer being realised to its full potential? Why or why not?
			\item Can you think of any improvements which would increase technology transfer?
		\end{itemize}
		9. *What was the impact (if any) on [X country] of joining the international intellectual property law systems under the TRIPS Agreement?
		\begin{itemize}
			\item Can you describe any advantages or disadvantages? Can you give any examples?
		\end{itemize}
		10. Do any aspects of IP law need to be reformed? Why or why not? What needs to change?\\
		11. I am sure you are aware of the debates surrounding IPRs in relation to COVID-19 vaccine technologies. What have we learned from these debates? 
		\begin{itemize}
			\item Do you think the COVID-19 pandemic and the climate emergency raise any similar issues? Why or why not?
		\end{itemize}
		12. Based on your view of climate-friendly innovation in [X country], would you say that IPRs promote climate-friendly innovation, or hinder it?
		\begin{itemize}
			\item For instance, would you agree with the statement that patents are a ‘barrier’ to green innovation, and if so why, or why not? What about other IPRs?
		\end{itemize}
		13. Is there anything else I didn't touch on you would like to cover?\\
		
		Note. * Denotes possible follow-up questions depending on the interviewee's role.
	\end{document}